\documentclass[journal]{IEEEtran}
\pdfoutput=1
\usepackage{cite}
\usepackage{amsmath,amssymb,amsfonts}
\usepackage{algorithmic}
\usepackage{graphicx}
\usepackage{textcomp}
\usepackage{framed,multirow}
\usepackage{latexsym}
\usepackage{amsmath}
\usepackage{threeparttable}
\usepackage{float}
\usepackage{array}
\usepackage{hhline}
\usepackage{colortbl}
\usepackage{stfloats}
\usepackage{booktabs}
\usepackage{multirow}
\usepackage{enumerate}
\usepackage{diagbox}
\usepackage{url}
\usepackage{hyperref}
\usepackage{makecell,rotating}

\def\etal{\textit{et~al}.}

\DeclareGraphicsExtensions{.pdf,.jpeg,.png,.jpg}

\graphicspath{{fig/}}

\definecolor{mygray}{gray}{.9}

\definecolor{newcolor}{rgb}{.8,.349,.1}

\makeatletter

\newcommand{\Rmnum}[1]{\expandafter\@slowromancap\romannumeral #1@}
\makeatother

\title{WSSS4LUAD: Grand Challenge on Weakly-supervised Tissue Semantic Segmentation for Lung Adenocarcinoma}

\author{Chu Han$^\dagger$, Xipeng Pan$^\dagger$, Lixu Yan$^\dagger$, Huan Lin$^\dagger$, Bingbing Li, Su Yao, Shanshan Lv, Zhenwei Shi, Jinhai Mai, Jiatai Lin, Bingchao Zhao, Zeyan Xu, Zhizhen Wang, Yumeng Wang, Yuan Zhang, Huihui Wang, Chao Zhu, Chunhui Lin, Lijian Mao, Min Wu, Luwen Duan, Jingsong Zhu, Dong Hu, Zijie Fang, Yang Chen, Yongbing Zhang, Yi Li, Yiwen Zou, Yiduo Yu, Xiaomeng Li, Haiming Li, Yanfen Cui, Guoqiang Han, Yan Xu, Jun Xu, Huihua Yang, Chunming Li, Zhenbing Liu, Cheng Lu, Xin Chen$^*$, Changhong Liang$^*$, Qingling Zhang$^*$, Zaiyi Liu$^*$
\thanks{Chu Han, Xipeng Pan, Huan Lin, Zhenwei Shi, Jinhai Mai, Bingchao Zhao, Zeyan Xu, Chao Zhu, Huihui Wang, Yuan Zhang, Haiming Li, Yanfen Cui, Cheng Lu, Changhong Liang and Zaiyi Liu are with the Department of Radiology, Guangdong Provincial People's Hospital, Guangdong Academy of Medical Sciences, Guangzhou, 510080 China; Guangdong Provincial Key Laboratory of Artificial Intelligence in Medical Image Analysis and Application, Guangdong Provincial People’s Hospital, Guangdong Academy of Medical Sciences, Guangzhou, 510080 China}
\thanks{Qingling Zhang, Lixu Yan, Su Yao and Shanshan Lv are with the Department of Pathology, Guangdong Provincial People’s Hospital, Guangdong Academy of Medical Sciences, Guangzhou, 510080 China}
\thanks{Zhenbing Liu, Zhizhen Wang and Yumeng Wang are with the School of Computer Science and Information Security, Guilin University of Electronic Technology, Guangxi, 541004 China}
\thanks{Bingbing Li is with the Department of Pathology, Guangdong Provicial People's Hospital Ganzhou Hospital, Ganzhou, China}
\thanks{Huihua Yang is with the College of Artificial Intelligence, Beijing University of Posts and Telecommunications, Beijing, 100876 China}
\thanks{Jiatai Lin and Guoqiang Han are with the School of Computer Science and Engineering, South China University of Technology, Guangzhou, 510640 China}
\thanks{Jun Xu is with the Institute for AI in Medicine, School of Artificial Intelligence, Nanjing University of Information Science and Technology, Nanjing, 210044 China}
\thanks{Yan Xu is with the School of Biological Science and Medical Engineering, State Key Laboratory of Software Development Environment, Key Laboratory of Biomechanics and Mechanobiology of Ministry of Education, Research Institute of Beihang University in Shenzhen, Beijing Advanced Innovation Center for Biomedical Engineering, Beihang University, Beijing, 100191 China}
\thanks{Xin Chen is with the Department of Radiology, Guangzhou First People's Hospital, the Second Affiliated Hospital of South China University of Technology, Guangzhou, 510180 China}
\thanks{Chunhui Lin, Lijian Mao and Min Wu are with the Zhejiang Dahua Technology Co. Ltd, Zhejiang, 310053 China}
\thanks{Luwen Duan is with the School of Biomedical Engineering, University of Science and Technology of China, Hefei, 230022 China}
\thanks{Zijie Fang, Yang Chen and Yongbing Zhang are with the Tsinghua Shenzhen International Graduate School, Tsinghua University, Shenzhen, China}
\thanks{Yi Li, Yiwen Zou, Yiduo Yu and Xiaomeng Li are with Electronic and Computer Engineering \& Computer Science and Engineering, The Hong Kong University of Science and Technology, Hong Kong}
\thanks{Jingsong Zhu is the School of Automation Science and Engineering, Xi'an Jiaotong University, Xian, 710049 China}
\thanks{Dong Hu is with Department of Computer Science, Huaqiao University, Xiamen, 361021 China}
\thanks{$^\dagger$Equal contribution}
\thanks{$^*$ Corresponding authors}
}

\begin{document}
\maketitle

\IEEEtitleabstractindextext{\begin{abstract}
Lung cancer is the leading cause of cancer death worldwide, and adenocarcinoma (LUAD) is the most common subtype. Exploiting the potential value of the histopathology images can promote precision medicine in oncology. Tissue segmentation is the basic upstream task of histopathology image analysis. Existing deep learning models have achieved superior segmentation performance but require sufficient pixel-level annotations, which is time-consuming and expensive. To enrich the label resources of LUAD and to alleviate the annotation efforts, we organize this challenge WSSS4LUAD to call for the outstanding weakly-supervised semantic segmentation (WSSS) techniques for histopathology images of LUAD. Participants have to design the algorithm to segment tumor epithelial, tumor-associated stroma and normal tissue with only patch-level labels. This challenge includes 10,091 patch-level annotations (the training set) and over 130 million labeled pixels (the validation and test sets), from 87 WSIs (67 from GDPH, 20 from TCGA). All the labels were generated by a pathologist-in-the-loop pipeline with the help of AI models and checked by the label review board. Among 532 registrations, 28 teams submitted the results in the test phase with over 1,000 submissions. Finally, the first place team achieved mIoU of 0.8413 (tumor: 0.8389, stroma: 0.7931, normal: 0.8919). According to the technical reports of the top-tier teams, CAM is still the most popular approach in WSSS. Cutmix data augmentation has been widely adopted to generate more reliable samples. With the success of this challenge, we believe that WSSS approaches with patch-level annotations can be a complement to the traditional pixel annotations while reducing the annotation efforts. The entire dataset has been released to encourage more researches on computational pathology in LUAD and more novel WSSS techniques.
\end{abstract}
}

\maketitle
\IEEEdisplaynontitleabstractindextext

\IEEEpeerreviewmaketitle

\section{Introduction}
Lung cancer is the leading cause of cancer death worldwide~\cite{sung2021global}, and lung adenocarcinoma (LUAD) is the most common subtype~\cite{Alesha2021lung}. During the diagnosis process of LUAD, histopathology examination of the tissue section, stained with hematoxylin and eosin (H\&E), remains the gold standard. It delivers massive information on tumor microenvironment from different perspectives, such as the morphology of the tissues, spatial arrange at tissue-level or cell-level, nuclear atypia and etc~\cite{hanahan2022hallmarks}. Experienced pathologists are able to evaluate the tumor progression by reading the patterns of the interaction between the tumor and its microenvironment~\cite{bremnes2011role}. But it is experience-dependent with limited interobserver agreement~\cite{yu2018artificial}. For an objective, quantitative and precise analysis of histopathology sections, it is urged to translate the visual language to the computer language.
\begin{figure}[t]
	\centering
	\includegraphics[width=.99\linewidth]{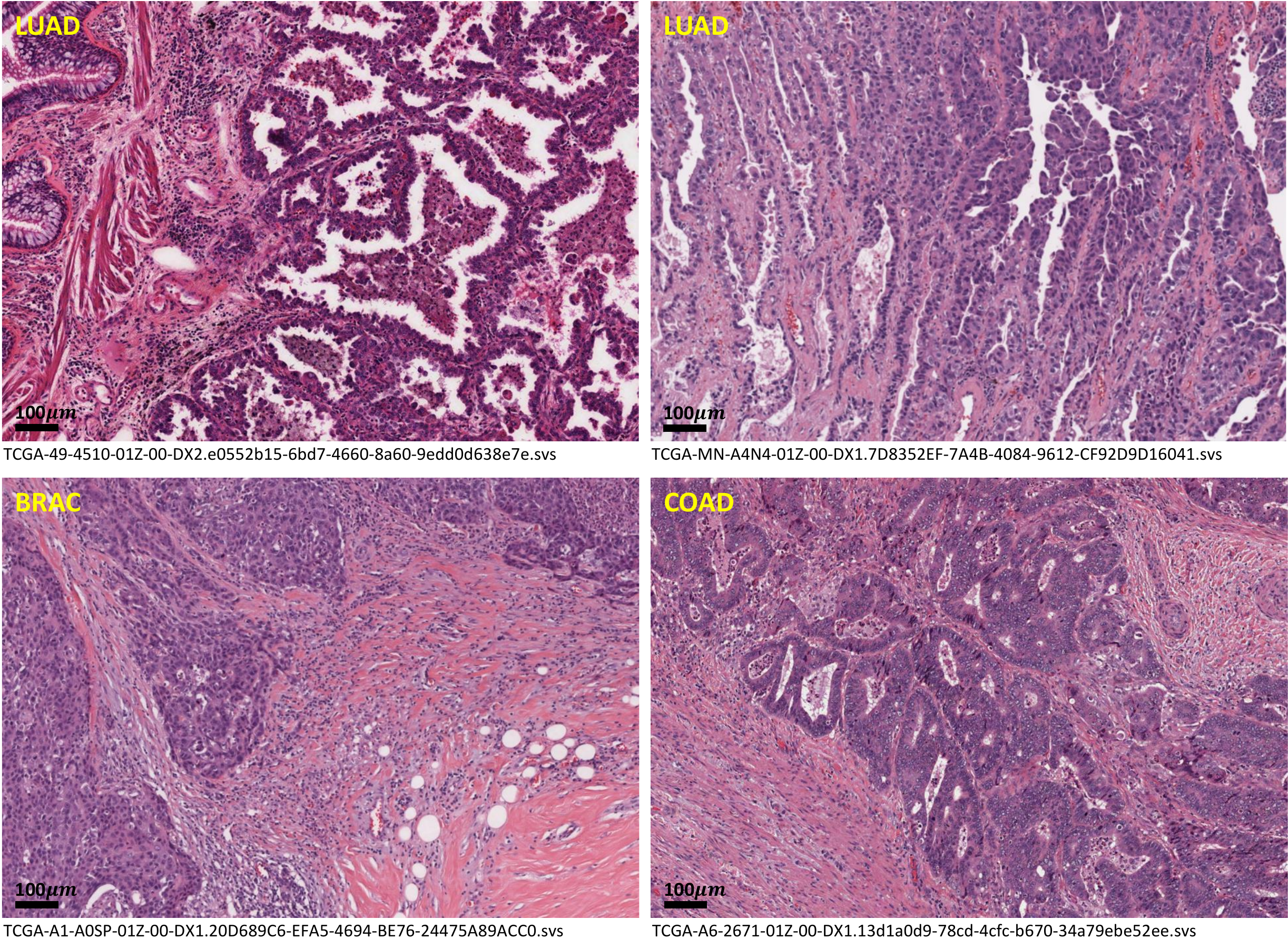}
	\caption{Tumor heterogeneity in different cancers.}
	\label{fig:tumor-heter}
\end{figure}

Tissue semantic segmentation is one of the most important upstream tasks in \textit{Digital Pathology}. It defines the tissue type for every single pixel and allows quantitative analysis for the downstream cancer outcome predictions~\cite{srinidhi2021deep}. The evolution of deep learning algorithms~\cite{lecun2015deep} is leading the image segmentation and recognition tasks to a computational and data-driven manner. These learning-based models achieved outstanding performance compared with traditional approaches by strong capacity neural network models~\cite{he2016deep,chen2017rethinking} under the supervision of a large among of annotated data~\cite{krizhevsky2012imagenet}. However, for the image segmentation task, drawing pixel-level annotation is extremely difficult and may spend a tremendous amount of time and resources~\cite{budd2021survey}. For histopathology images, such difficulties have been even amplified due to the specialty of the whole slide images (WSI). First of all, WSIs are scanned by microscopic slide scanners. The image resolution of WSIs can be up to over 100000$\times$100000. Drawing pixel-level annotations in such huge images are much harder than in natural images. Second, malignant tumors are heterogeneous. Tumor epithelial has complex morphological appearances, which leads to ambiguities between two tissue types. For lung adenocarcinoma, the labeling process is even harder than for other malignant tumors. As can be seen in Fig.~\ref{fig:tumor-heter}, the interactions between tumor and tumor-associated stroma in LUAD are twisted and more complex than other two cancers (from TCGA~\footnote{\url{https://www.cancer.gov/tcga}}). Last but not least, because of the image specialty, crowdsourcing by normal people with no medical background is incapable of the labeling job. It may spend a long time for pathologists to collect sufficient pixel-level training samples for tissue semantic segmentation models.

To tackle these problems, we set up the WSSS4LUAD challenge with two goals. 1) Introducing more public labeling resources of WSIs in lung adenocarcinoma, to promote the research on lung cancer histopathology image analysis. 2) Discovering a more efficient and easier way to accelerate the labeling process and alleviate the annotation efforts of pathologists for tissue semantic segmentation. In computer vision, some researchers are focusing on using image-level labels to achieve semantic segmentation, called weakly-supervised semantic segmentation (WSSS). In this challenge, participants have to build innovative \textbf{WSSS} techniques to achieve tissue semantic segmentation for \textbf{LU}ng \textbf{AD}enocarcinoma. That is the reason why this challenge is called WSSS4LUAD. To the best of our knowledge, this is the first challenge and dataset for WSSS in lung adenocarcinoma. We defined three majority tissue types for lung adenocarcinoma, including tumor epithelial, tumor-associated stroma and normal tissue. WSSS4LUAD was organized as a part of the ISICDM 2021 (the 5th International Symposium on Image Computing and Digital Medicine). With the outstanding WSSS algorithms tailored for histopathology images, we hope that the conventional way of drawing pixel-level annotations can be replaced by a simpler way of clicking and selecting the classes of the patch.
\section{Related Works}
In this paper, we briefly review the recent works on histopathology image analysis for lung cancer, histopathology image segmentation and annotation efficient models.
\subsection{Lung Cancer - Histopathology Image Analysis}
Histopathology slides are the golden standard for lung cancer diagnosis~\cite{THAI2021535}. However, only relying on the visual inspection of the pathologists cannot fully discover the potential value of the histopathology slides. The invention of whole slide image scanners and the development of artificial intelligence techniques bring computational pathology to lung cancer~\cite{wang2019artificial}. Comprehensive analysis in different levels of the whole slide images is now available with the computing power of GPU (graphics processing unit), including image-level feature analysis, tissue-level quantitative analysis and cell-level spatial distribution analysis~\cite{sakamoto2020narrative}. 

Yu~\etal~\cite{yu2016predicting} extracted a large set of image features and proposed a machine learning pipeline for risk-stratification in patients with non-small cell lung cancer (NSCLC). Coudray~\etal~\cite{coudray2018classification} applied an end-to-end CNN model for lung cancer diagnosis. The model classified lung cancer into three categories, adenocarcinoma (LUAD), squamous cell carcinoma (LUSC) and normal lung tissues. The diagnostic performance of the AI model was equivalent to that of pathologists with the area under the curve (AUC) of 0.97. The authors also demonstrated that deep learning models could be used to predict molecular status, like STK11, EGFR, FAT1, SETBP1, KRAS and TP53 mutant genes. Abduljabbar~\etal~\cite{abduljabbar2020geospatial} analyzed the spatial distribution of different types of cells and found heterogeneity of immune status, defined as two immune phenotypes (immune hot and immune cold). The immune phenotype can predict the risk of cancer recurrence of lung adenocarcinoma. Zheng~\etal~\cite{zheng2020spatial} discussed the interactions between tumor cells and tumor-associated macrophages (TAM) in multiple immunofluorescence staining images and proved the density and spatial distribution of TAM phenotypes to be the prognostic factors for lung cancer survival. Lu~\etal~\cite{lu2020prognostic} constructed interpretable features by analyzing tumor cellular diversity (CellDiv), which was proven to be a strong prognostic factor of 5-year overall survival in NSCLC patients validated by the multi-center data. Recently, an AI-powered spatial analysis model~\cite{JCO2022} has been proposed to analyze tumor-infiltrating lymphocytes in the tumor microenvironment, in order to find the correlation with immune checkpoint inhibition in advanced NSCLC. This research team has made around $2.8\times10^9 \mu \rm{m}^2$ tissue-level annotated area in over 3,000 tumor slides, which is a tremendous annotation workload.

Even a lot of excellent researches have been proposed for histopathology image analysis for lung cancer, researchers still have to spend much time collecting a large among of annotations for the segmentation model. Therefore, we organize this challenge to enrich the annotation resources and propose a simpler labeling way to relieve the annotation efforts.

\subsection{Fully-supervised Histopathology Image Segmentation}
The main contribution of computational pathology~\cite{srinidhi2021deep} is to automatically, precisely, objectively and quantitatively assess the whole slide image. Histopathology image segmentation is the most basic task in this field. It can help differentiate the constituents of the tumor in different levels, such as tissue-level~\cite{HookNet}, gland-level~\cite{chen2016dcan} and cell-level~\cite{zhao2020nuclei,graham2019hover}, translating the visual language to the computer language.

Recently, more and more tissue segmentation challenges have been held such as, BCSS~\cite{amgad2019structured} for breast cancer, Gleason 2019~\cite{nir2018automatic} for prostate cancer, DigestPath 2019~\cite{li2019signet} for colon cancer and LungHP Challenge 2019~\cite{li2020deep} for lung cancer. LungHP Challenge 2019 is the only one we found in this field. It aims to segment cancer tissue in WSI in three different lung cancer subtypes, SCC, SCLC and ADC. By exploring the digital pathology communities, we found that tissue-level label resources are still limited, especially in lung cancer. That is one of the reasons why we organize this challenge to promote and complement the histopathology image analysis in lung cancer.

However, the conventional manual annotation way is slow, inefficient and annoying. And this work cannot be done in a crowdsourcing way like natural images, because people with no medical background are not qualified for this. Therefore, we want to introduce computer algorithms in this process, pushing the labeling job to an annotation-efficient manner.

\subsection{Annotation Efficient Approaches}
With the great success of fully-supervised deep learning models, researchers now focus on solving medical image segmentation tasks in an annotation efficient manner. It can be roughly categorized into several directions, active learning, self-supervised learning and not-so-supervised learning.

\subsubsection{Active Learning}
Active learning (AL)~\cite{budd2021survey} is a popular technique in medical image segmentation due to the expertise requirements of the labelers. The concept of AL is to let AI models decide what should be labeled, which are commonly the most valuable and informative samples~\cite{wen2018comparison,shen2020deep}.

Doyle~\etal~\cite{Doyle2011} used the AL approach to select the most informative regions in prostate histopathology images. Such a strategy can alleviate the class-imbalance problem of the minority classes.
Zhou~\etal~\cite{zhou2021active} associated active learning with transfer learning to select the worthy samples while continuously improving the model by continual fine-tuning. 
Belharbi~\etal~\cite{belharbi2021deep} introduced a self-training model to the active learning process. They used the segmentation pseudo masks to guide the classification tasks.  
Greenwald~\etal~\cite{greenwald2021whole} proposed a pathologist-in-the-loop TissueNet to create the largest cell dataset for fluorescent multiplex immunohistochemistry with over a million labeled cells.

\subsubsection{Self-supervised Learning}
Self-supervised learning~\cite{shurrab2021self} aims to learn a good initialized feature representation for the upcoming task by a pretext task without the necessity of human annotations, including contrastive, predictive and generative tasks.

Contrastive learning~\cite{van2018representation} trains the model by measuring the similarity and difference between the images. Following this idea, various famous works have been proposed soon, including MoCo~\cite{he2020momentum}, SimCLR~\cite{chen2020simple} and BYOL~\cite{grill2020bootstrap,wang2021transpath}.
Predictive learning sets up some prediction tasks with the generated labels by the prior knowledge of the images, such as image reconstruction~(\cite{he2021masked}), jigsaw puzzle solving~\cite{li2020self} and Rubik cube solving~\cite{zhuang2019self,zhu2020rubik}.
Generative models learn the domain knowledge by restoring the manually corrupted images like image inpainting~\cite{pathak2016context}.

In histopathology image analysis, Srinidhi~\etal~\cite{srinidhi2022self} proposed a ``resolution sequence prediction task'' to mimic how pathologists read the WSIs at different magnifications.
Boyd~\etal~\cite{boyd2021self} introduced a visual field expansion to use the input image to generate the pseudo contents of the surrounding area.
Cheng~\etal~\cite{cheng2021curriculum} proposed a curriculum self-supervised learning strategy by training several tasks with progressively increasing difficulties. It includes the image reconstruction task, the image inpainting task and the stain deconvolution task. 
Luo~\etal~\cite{luo2022sdmae} proposed an SD-MAE by associating self-distillation augmented SSL with masked autoencoder and improved the classification performance of histopathology images.
Numerous approaches have already proven that the self-supervised pre-training can effectively learn the domain-specific feature representations and the inherent image characteristics to alleviate the training burden and the data requirement for the upcoming tasks.

\subsubsection{Not-so-supervised Learning}
Besides training a well-initialized model by self-supervised pretraining approaches, researchers also try their best to reduce the annotation efforts in a not-so-supervised manner~(\cite{cheplygina2019not}), such as to simplify the problem by patch-level annotations (patch-level classification), to maximize the value of the scarce labels (semi-supervised learning) or to replace pixel-level labels by weaker labels (weakly-supervised learning).


\textbf{Patch-level classification:}
To alleviate dense pixel-level annotation for tissue segmentation, Kather~\etal~\cite{Kather2019,kather2019predicting} simplified the semantic segmentation problem to a patch-level classification problem, which scarified the boundaries precision. Following this idea, Lin~\etal~\cite{lin2022pdbl} proposed a deep-broad plug-and-play module (Pyramidal Deep-Broad Learning, PDBL) that can be easily adapted to any well-trained CNN classification backbone. The proposed PDBL can greatly reduce the training burden and annotations while maintaining a promising classification accuracy. 

\textbf{Semi-supervised learning:}
Semi-supervised learning explores the semantic connections between the labeled images and the unlabeled images. Mainstream models are usually formulated according to the following two spirits, generating pseudo labels~\cite{chai2021deep} or maintaining the semantic consistency~\cite{xie2020pairwise,marini2021semi}. They are based on the assumption that the distribution between labeled and unlabeled images should be close enough, which is an obstacle in histopathology images. Due to the heterogeneous tumor microenvironment, the same tissue class may show totally different morphological appearances. For histopathology image segmentation, weaker labels with larger sample varieties in a weakly-supervised manner may be more effective than dense annotations with limited sample varieties in a semi-supervised manner.

\textbf{Weakly-supervised learning:}
Researchers also tried different kinds of weak labels to replace dense pixel ones.
Multiple instance learning models~\cite{jia2017constrained,lerousseau2020weakly} were proposed for cancerous region segmentation given the slide-level labels only. But it can only segment cancerous or non-cancerous regions.
Scribble-based models were introduced to perform a fast user interaction and label adjustment in both cellular-level~\cite{lee2020scribble2label} and tissue-level~\cite{jahanifar2021robust}.
Various point-based models~\cite{qu2019weakly,qu2020weakly,koohbanani2020nuclick,lin2022label} have proven the effectiveness on the nuclei segmentation task.
Some recent works~\cite{Pseudo_Labeling,ZHANG2021102183} utilized the proportion of the tissue as the labels. 

CAM-based models~\cite{zhou2016learning} have been widely used in weakly-supervised semantic segmentation (WSSS), including histopathology images~\cite{chan2019histosegnet}. The organizers of WSSS4LUAD challenge also proposed a CAM-based model~\cite{han2021multi} to achieve WSSS for tissue segmentation. They introduced a progressive dropout attention mechanism to gradually deactivate the most distinguishable regions. A multi-layer pseudo-supervision was proposed to enrich the information. In this challenge, we apply this model to generate the initial pixel-level pseudo masks for the validation and test sets, which greatly saves the annotation times of the pathologists.

\section{Challenge Description}
\begin{figure*}[htp]
	\centering
	\includegraphics[width=.99\linewidth]{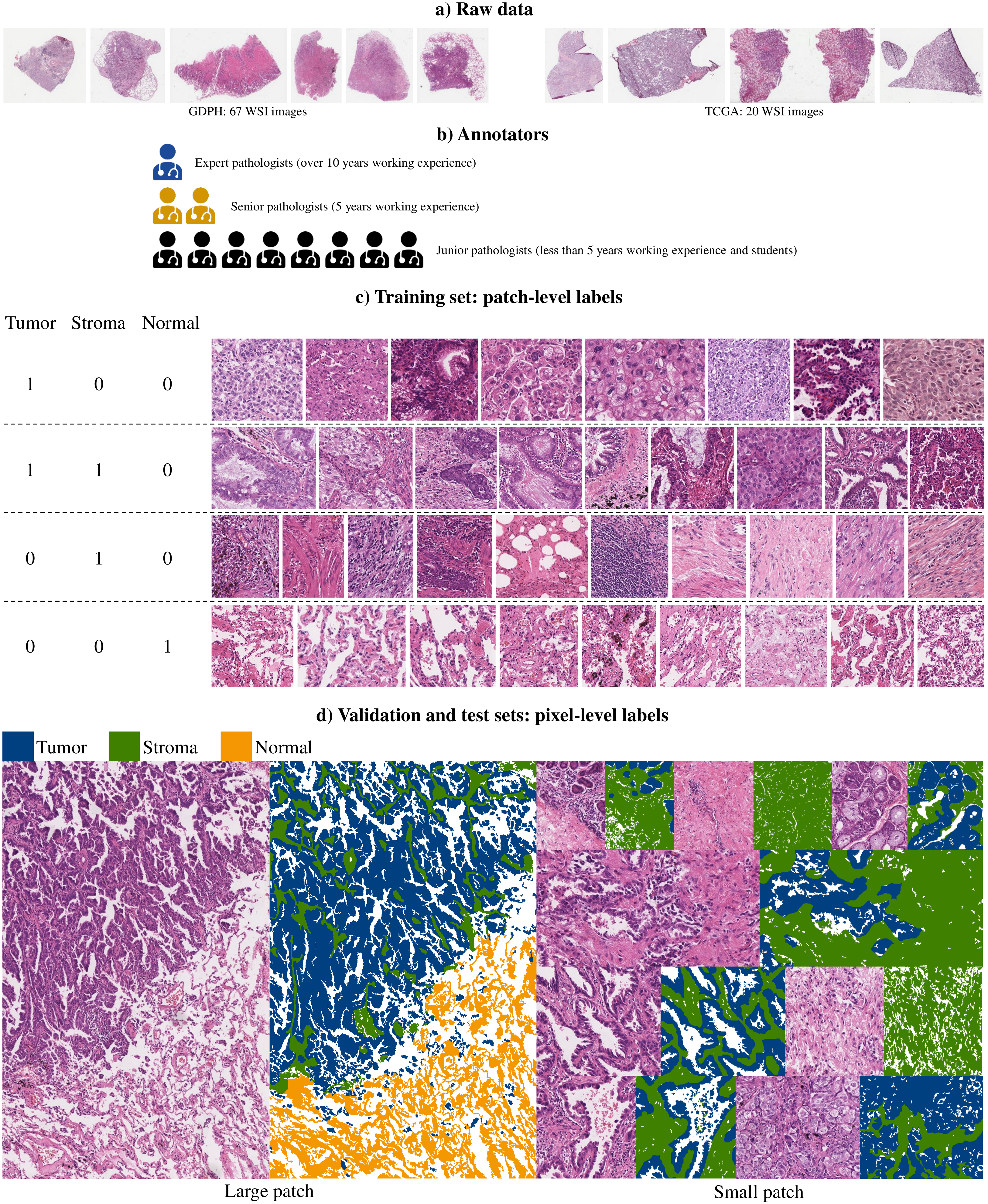}
	\caption{Brief introduction of WSSS4LUAD challenge. (a) We scanned 67 histopathology slides (one slide per patient) from Guangdong Provincial People's Hospital and downloaded 20 WSIs from TCGA. (b) one expert, two senior and 8 junior pathologists were invited to annotate image-level training samples and pixel-level validation samples. (c) and (d) shows some examples of the training, validation and test sets.}
	\label{fig:pipeline}
\end{figure*}

In this challenge, we aim to use patch-level classification labels to predict pixel-level semantic segmentation maps with three tissue categories, including tumor epithelial (tumor), tumor-associated stroma (stroma) and normal tissue (normal). So we prepare patch-level classification labels for the model training and dense pixel-level annotations to evaluate the model performance. A brief description of this challenge is shown in Fig.~\ref{fig:pipeline}. 
\begin{figure*}[t]
	\label{fig:label_pipeline}
	\centering
	\includegraphics[width=.99\linewidth]{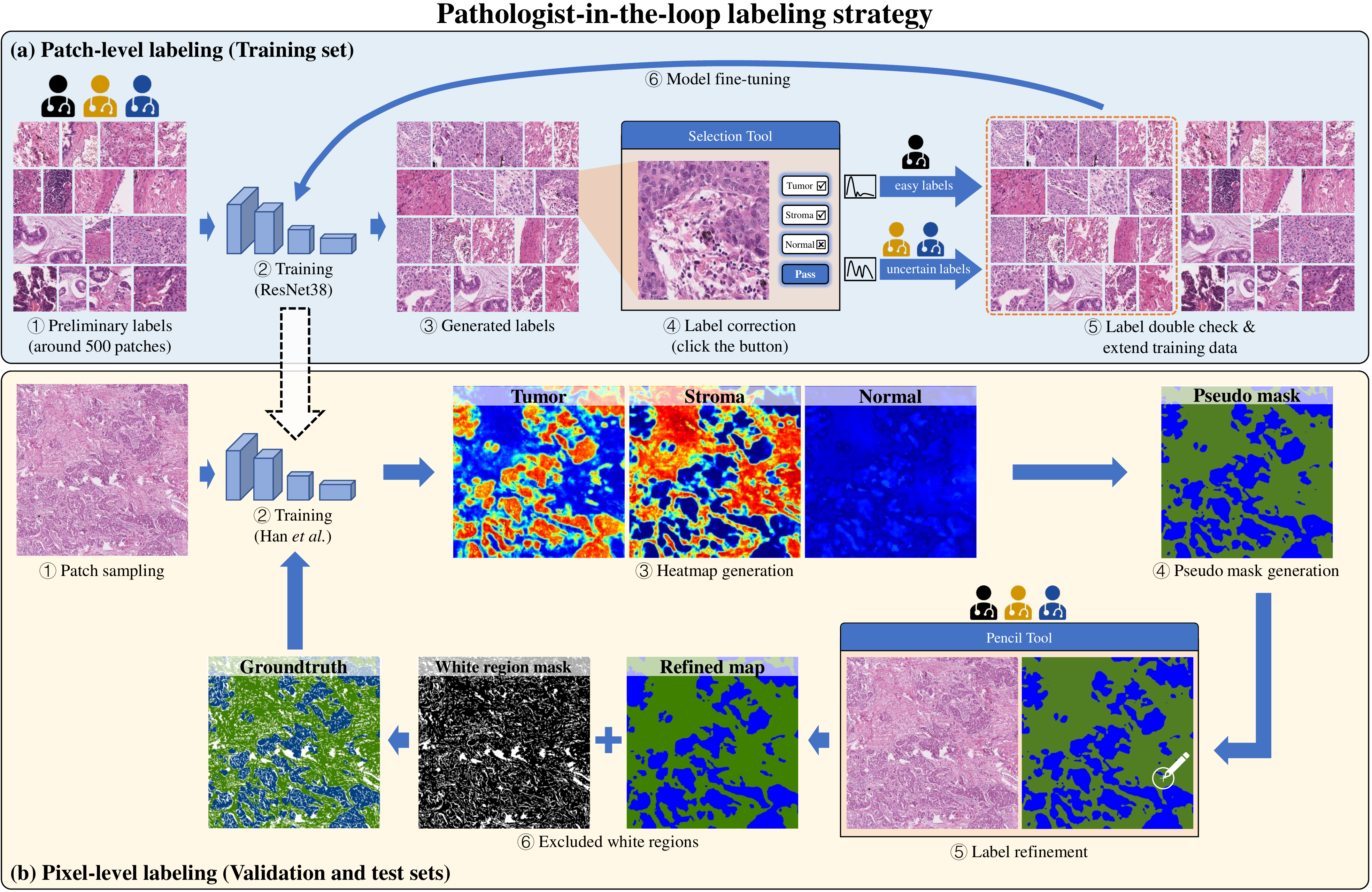}
	\caption{The pipeline of the pathologist-in-the-loop labeling strategy. (a) Patch-level labeling for the training set. Pathologists first annotate 500 initial patch labels to train a preliminary model (ResNet38) and generate a set of pseudo labels. Then they use a selection tool to check and correct the pseudo labels. Pseudo labels with high confidence are verified by junior pathologists. The ones with high uncertainty are corrected by senior pathologists. Verified labels are then fed to fine-tune the classification model. (b) Pixel-level labeling for the validation and test sets. The well-trained ResNet38 in (a) is inherited in this phase to generate pixel-level semantic segmentation maps by our previous work MLPS~\cite{han2021multi}. Pathologists refine the pseudo masks using the pencil tool of Adobe Photoshop. The verified ground truths are added to the training set to fine-tune the model. *For both tasks, with more corrected labels, the model improves and the manual intervention decreases.}
\end{figure*}

\subsection{Whole Slide Images Preparation}
The whole slide images are from two sites, including the Department of Pathology, Guangdong Provincial People's Hospital and TCGA (multiple centers). Since LUAD is a highly heterogeneous tumor, in order to ensure the diversity of the tissue-level morphological appearance, we scan 67 H\&E stained histopathology sections from 67 LUAD patients with surgical resection and download 20 WSIs from TCGA. All the slides are scanned by a microscopic digital slide scanner (Leica GT 450) at a 40X magnification with 0.2517 µm/pixel resolution and are under quality control by filtering out the blurry, over-stained, under-stained and dirty slides. We do not perform stain normalization to keep the unharmed source images for the participants. Because of the ethical restrictions, the WSIs in our center cannot be released so far. We are applying for approval from the ethical committee. Hopefully, the source images can also be released in the future.

\subsection{Annotators Recruitment and Label Review Board}
As demonstrated in Fig.~\ref{fig:pipeline} (b), we invite 11 annotators with medical backgrounds to label the data and to verify the annotations, including one expert pathologist with over 10 years of working experience in lung cancer diagnosis, two senior pathologists with 5 years of working experience, and 8 junior pathologists with less than 5 years working experience or medical students. All the labeling jobs are assigned to junior pathologists. The expert pathologist and two senior pathologists form a label review board to verify the labels. We introduce three mechanisms to guarantee consistency among different labelers. 1) Before we start the labeling job, the expert pathologist is asked to give a talk about the labeling criteria. 2) At the beginning of the labeling process, we ask all the labelers to annotate a few samples. And then all the labels are reviewed by the label review board and form documentation of the label suggestions, which will be returned to each labeler to adjust their labeling criteria. 3) We introduce a certain degree (around 50\%) of overlapping samples among adjacent labeling assignments.

\begin{figure*}[t]
	\centering
	\includegraphics[width=.99\linewidth]{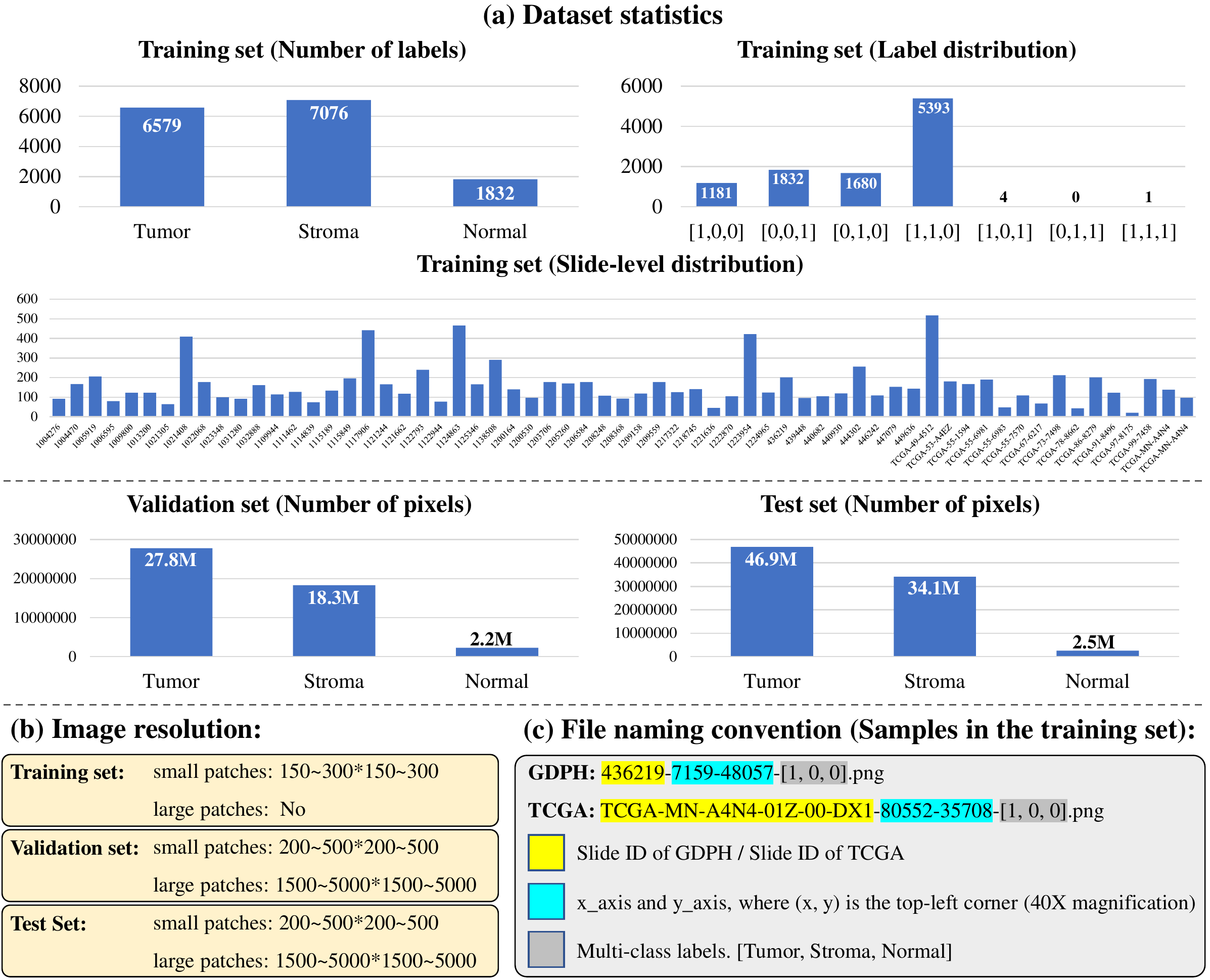}
	\caption{More details of the dataset. (a) Data statistics include the number of the patch-level and pixel-level labels for each tissue category in the training, validation and test sets, respectively. And the distribution of the multi-class labels in the training set. (b) demonstrates the image resolution of the dataset. (c) shows the file naming convention of the sample in the training set.}
	\label{fig:data_statistics}
\end{figure*}

\subsection{Pathologist-in-the-loop Labeling Process}
There are two kinds of labels we prepare for this challenge, patch-level and pixel-level. To accelerate the annotation process and alleviate the annotation efforts, we introduce a pathologist-in-the-loop labeling manner for both annotation levels, demonstrated in Fig.~\ref{fig:label_pipeline}. Such a process can greatly accelerate the annotation process and reduce the annotation efforts.

\textbf{Patch-level annotations}: The way we construct the training set is demonstrated in Fig.~\ref{fig:label_pipeline}. First, pathologists have to annotate a set of classification labels (around 500 patches) as the initial training set. Then we train a preliminary classification model of ResNet38 to generate pseudo labels. The reason why we choose ResNet38 instead of more popular classification backbones (like ResNet50 or VGG16) is that ResNet38 is one of the most common backbones for weakly-supervised semantic segmentation~\cite{han2021multi}. And the trained ResNet38 model will be used to generate pixel-level labels in the next step. Next, pathologists use a selection tool to check and correct the pseudo labels. During the label correction process, we stratify the pseudo labels by the model predictive confidence. Pseudo labels with high confidence are verified by junior pathologists. The ones with low confidence and high uncertainty are corrected by senior pathologists. All the patch labels are verified by at least two pathologists. The inconsistent labels are further checked by the label review board. Finally, the verified labels are then fed to fine-tune the classification model.

\textbf{Pixel-level annotations}: We introduce our previous proposed approach MLPS~\cite{han2021multi}, a weakly-supervised semantic segmentation model tailored for the histopathology images. We use the patch-level labels collected in the previous phase to train MLPS. This model generates the heatmap by the class activation maps technique for each tissue category. Then we obtain the pseudo masks by combining the heatmaps. Pathologists refine the pseudo masks using the pencil tool of Adobe Photoshop. The verified ground truths are added to the training set to fine-tune the model. Note that, the large patches (over 1000$\times$1000 resolution) are first cut into several image tiles, and then stitched back to the original size. All the pixel-level labels are finally reviewed and confirmed by the label review board.

\subsection{Dataset Details}
Here, we show the details of the dataset, including the statistics, image resolution and file naming convention, shown in Fig.~\ref{fig:data_statistics}. Note that, the labels of the training samples are three digit classification labels, [Tumor, Stroma, Normal]. All the patches are under 10$\times$ magnification.

\subsubsection{Training set}
In the training set, we randomly sample 49 WSIs from GDPH and 14 WSIs from TCGA.
A total of 15,000 patches are randomly cropped from the WSIs with arbitrary resolution (around 150-300*150-300). During the histopathology slide-making procedure, dust may exist in the slides inevitably, which will cause out-of-focus blurs in some local regions when scanning the slides. Therefore, all the patches are under quality control to filter out the blurry and dirty regions. Furthermore, some highly ambiguous patches are also removed to avoid introducing uncertain information to the model. Finally, 10,091 patches are left in the training set.

The statistics of the training data are shown in Fig.~\ref{fig:data_statistics} (a). As can be seen, the largest number of labels are [1,1,0], which is tumor and stroma without normal. The second most labels are [1,0,0], [0,0,1] and [0,1,0] with only one tissue categories. There are some label combinations are very few even is nonexistent. Because of the invasiveness of LUAD, it is hard to observe a clear boundary of the tumor and normal tissue. Therefore, it is almost impossible to cover normal and tumor/stroma within such a small patch.

We put all the useful information of the training samples into the filename, including the slide ID, $(x,y)$ coordinates and classification labels, shown in Fig.~\ref{fig:data_statistics} (c).

\subsubsection{Validation and test sets}
We select 18 WSIs from GDPH and 6 WSIs from TCGA. Then the WSIs are randomly and equally separated to the validation set (GDPH: 9, TCGA: 3) and the test set (GDPH: 9, TCGA: 3). Total 40 patches are manually cropped by the label review board for the validation set, including 9 large patches (around 1500-5000*1500-5000) and 31 small patches (around 200-500*200-500). Total 80 patches are manually cropped by the label review board for the test set, including 14 large patches (around 1500-5000*1500-5000) and 66 small patches (around 200-500*200-500). Fig.~\ref{fig:data_statistics} (a) shows the total number of labeled pixels in the validation and test sets. The ground truth pixel-level labels of each tissue category are defined as follows.

\begin{itemize}
	\item Tumor epithelial: (0, 64, 128), (label: 0)
	\item Tumor-associated stroma: (64, 128, 0), (label: 1)
	\item Normal: (243, 152, 0), (label: 2)
	\item White background: (255, 255, 255), (label: 3)
\end{itemize}

Since lungs are the main organ of the respiratory system. There are a lot of alveoli (some air sacs) serving for exchanging the oxygen and carbon dioxide, which forms some white background in WSIs. In this challenge, we pre-define the white background by a simply threshold and exclude it when evaluating the model.

\subsection{Evaluation Metrics}
In this challenge, we use mIoU to evaluate the models. The white backgrounds inside the alveoli are excluded when calculating mIoU. We have also provided a background mask for each patch in the validation and test sets.

Participants have to only use the provided training data to train their own models. External histopathological data is not allowed in this challenge. ImageNet pre-trained model is allowed as the initialization.
\section{Challenge Summary}
\begin{figure}[htp]
	\centering
	\includegraphics[width=.99\linewidth]{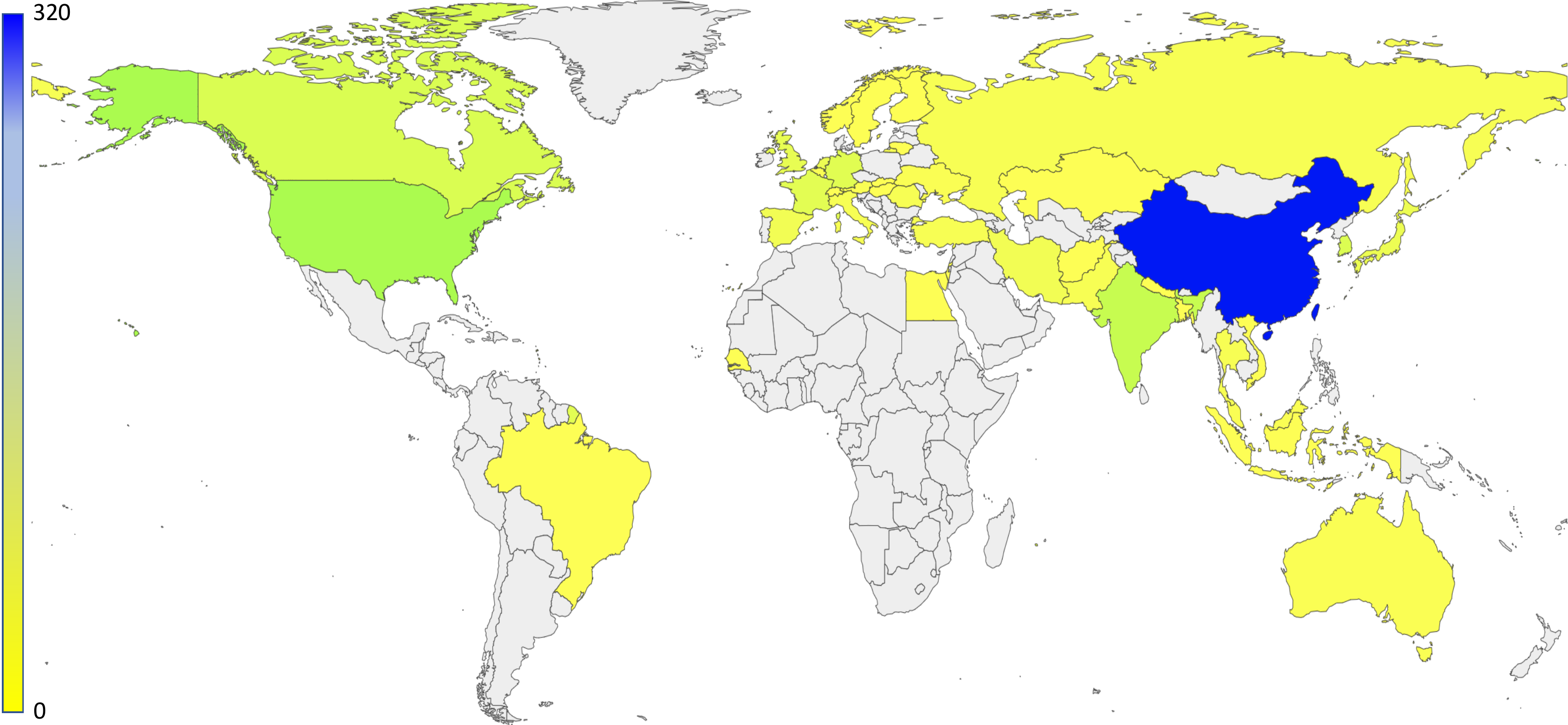}
	\caption{The locations of the registrations. China 299, United States of America 36, India 23, Canada 15, South Korea 14, Germany 12, France 11, Hong Kong 10, United Kingdom 9, Taiwan 8, Netherlands 7, Japan 7, Iran 5, Spain 5, Singapore 4, Russia 3, Romania 3, Italy 3, Pakistan 3, Turkey 3, Thailand 3, Ukraine 2, Belgium 2, Australia 2, Austria 2, Malaysia 2, Egypt 1, Finland 1, Sweden 1, Brazil 1, Norway 1, Indonesia 1, Israel 1, Bangladesh 1, Switzerland 1, Senegal 1, Hungary 1, Macao 1, Slovakia 1, Nepal 1, Vietnam 1, Kazakhstan 1, Afghanistan 1, Lithuania 1, unknown 20.}
	\label{fig:map}
\end{figure}
In this section, we summarize the challenge in the following aspects, including the registrations and challenge results (quantitative and qualitative). 

\subsection{Brief Summary}
Until now, there are 532 participants have registered for this challenge to access the data. Fig.~\ref{fig:map} demonstrates the locations of the registration. The top five countries or areas are China, USA, India, Canada and South Korea. 28 teams submit the results in the test phase with over 1000 submissions. Among them, 14 teams achieved over 0.7 mIoU in the test phase, as shown in Table~\ref{tab:qualitative}. Top-3 teams are required to submit the code, model and the introduction of the methodology. For the other teams, they are optional. Next, we demonstrate the methodology of the top-tier teams. 

\subsection{Top-tier Methods}\label{sec:technical}
\subsubsection{Lin~\etal~(1st place, Team: ChunhuiLin)}
Lin~\etal~propose a Siamese multi-task learning model for this challenge with a classification task and a segmentation task. CAM generated by the classification task is regarded as the pseudo-mask of the segmentation task. ASPP is applied to obtain a larger receptive field. They introduce a cross-supervision strategy between two identical siamese models. Here are the details of the proposed model. 
 
\textbf{Pre-processing:} 
Since the histopathology images are different from natural images, Lin~\etal~first calculate the mean and variance of the dataset to replace the parameters of ImageNet. Then, they upsample the number of normal tissue to re-balance the class by data augmentation. For the patches with only one class, they generate fake images by cutmix data augmentation to provide the pixel-level ground truth for the segmentation task. 

\textbf{Model:}
The same as the most conventional weakly-supervised semantic segmentation method, Lin~\etal~generate pseudo masks by calculating the class activation maps via a multi-label image classification model. They find that half of the training patches are single-class and the background of the histopathology image is relatively simple. Thus, Lin~\etal~use traditional image processing methods to generate more patch-level pseudo labels to enhance the classification model. After that, Lin~\etal~introduce an additional semantic segmentation branch and form a multi-task learning model. Two tasks share the same encoder and use different decoders. At the same time, they construct two datasets for each task. One is more than 4000 images with pixel-level pseudo labels, and the other one contains image-level labels. Based on the multi-task learning model, they construct a siamese multi-task network. Two multi-task learning models with identical structures. A cross-supervision strategy is proposed to use the pseudo masks generated by siamese models to supervise each other.

\textbf{Training \& inference:}
Lin~\etal~improve the loss function through joint Lovasz loss and cross-entropy loss. Online hard negative mining is applied during training. During the inference, they ensemble the outputs of the Siamese model. In addition, model ensemble has been done in this challenge, including ResNet101 and ResNeSt269 based on Deeplabv3 and HRNetw48.

\textbf{Post-processing:}
Lin~\etal~applied the BFS algorithm to fill the false background pixel by searching the nearest neighbor non-background pixel. In cancer pathological sections, Tumor and Stroma are often intertwined, but normal tissues hardly overlap with these areas. Based on this prior knowledge, they calculate the distribution of each category in the neighborhood (100x100) of each Tumor and Stroma pixel. After that, they modify the Tumor or Stroma pixels that accounted for more than 30\% of the normal category in the neighborhood to normal pixels.

\subsubsection{Zhu~\etal~(2nd place, Team: baseline0412)}
The task is the pixel-level prediction for tumor epithelial tissue, tumor stromal tissue and normal tissue. Since only image-level annotations (3-digit labels) are available in the training set, Zhu~\etal~employed weakly-supervised learning model.

\textbf{Dataset Enhancement:}
The concept of image splicing is to use parts of two or more images to create a new fake image. In this challenge, Zhu~\etal~apply the image splicing method to enhance the dataset in several aspects. First, they select the training samples with a single class, which only contains tumor
epithelial, tumor-associated stroma or normal tissue, and extract the foreground
region by using threshold segmentation. These single-class patches are used as ground truth samples for the segmentation task. Then, for the above single-class patches, they further apply the image splicing method to generate more fake image and ground truth pairs. They achieve this by randomly cutting a small piece from one patch and pasting it on the patch with different classes. And the patches with multiple classes are generated. Next, Zhu~\etal~also apply the image splicing method to the pseudo masks generated by CAM.

\textbf{Model:}
Weakly supervised models often use deep neural networks to generate the Class Activation Mapping (CAM), and to serve as pseudo-labels for the segmentation network. However, only relying on CAMs cannot obtain precise region boundaries. In order to enhance the localization accuracy of class activation mapping, they propose a multi-task learning model with the image splicing method. The proposed model consists of two modules, the first module is the classification model with CAM which is commonly used in weakly supervised semantic segmentation. The second module is a segmentation network guided by pseudo masks.

\begin{table}
	\centering
	\label{tab:qualitative}
	\caption{Qualitative results of the leaderboard (Test phase).}
	\begin{tabular}{c|cccc}
		\toprule
		Team name & mIoU & Tumor & Normal & Stroma\\
		\hline
		ChunhuiLin 		& \textbf{0.8413} & 0.8389 & \textbf{0.8919} & \textbf{0.7931}\\
		baseline0412	& 0.8222 & \textbf{0.8402} & 0.8343 & 0.7921\\
		Vison307		& 0.8058 & 0.8165 & 0.8554 & 0.7456\\
		BinghongWu		& 0.8057 & 0.8045 & 0.8654 & 0.7471\\
		adbertyoungdalu	& 0.8025 & 0.7967 & 0.8668 & 0.7440\\
		DPPD			& 0.7815 & 0.7895 & 0.8397 & 0.7153\\
		chenxl			& 0.7714 & 0.7897 & 0.8159 & 0.7088\\
		sibet0222		& 0.7609 & 0.8121 & 0.7107 & 0.7599\\
		guoxutao		& 0.7552 & 0.8179 & 0.6840 & 0.7636\\
		shichuanyexi	& 0.7411 & 0.8192 & 0.6714 & 0.7325\\
		zyw19990916		& 0.7382 & 0.8080 & 0.6868 & 0.7196\\
		York			& 0.7239 & 0.8023 & 0.6710 & 0.6985\\
		akiliyiu@gmail.com	& 0.7199 & 0.7557 & 0.7248 & 0.6791\\
		Zlin3000		& 0.7064 & 0.7493 & 0.6863 & 0.6837\\
		\toprule
	\end{tabular}
\end{table}

\subsubsection{Fang~\etal~(3rd place, Team: Vison307)}
\textbf{Pseudo Images and Masks Generation:}
The same as the top two teams, Fang~\etal~also perform a pseudo image generation by the patches with the single class. They randomly select patches with labels [1, 0, 0], [0, 1, 0] and [0, 0, 1] and randomly crop the patches and concatenate them to a new patch. Such a process can generate a set of multi-label patches with pixel-level pseudo masks to pre-train a segmentation model. Then, they augment the data by introducing rotations and flips. Cutmix strategy is also applied to further increase the training samples.

\begin{figure*}[t]
	\centering
	\includegraphics[width=.99\linewidth]{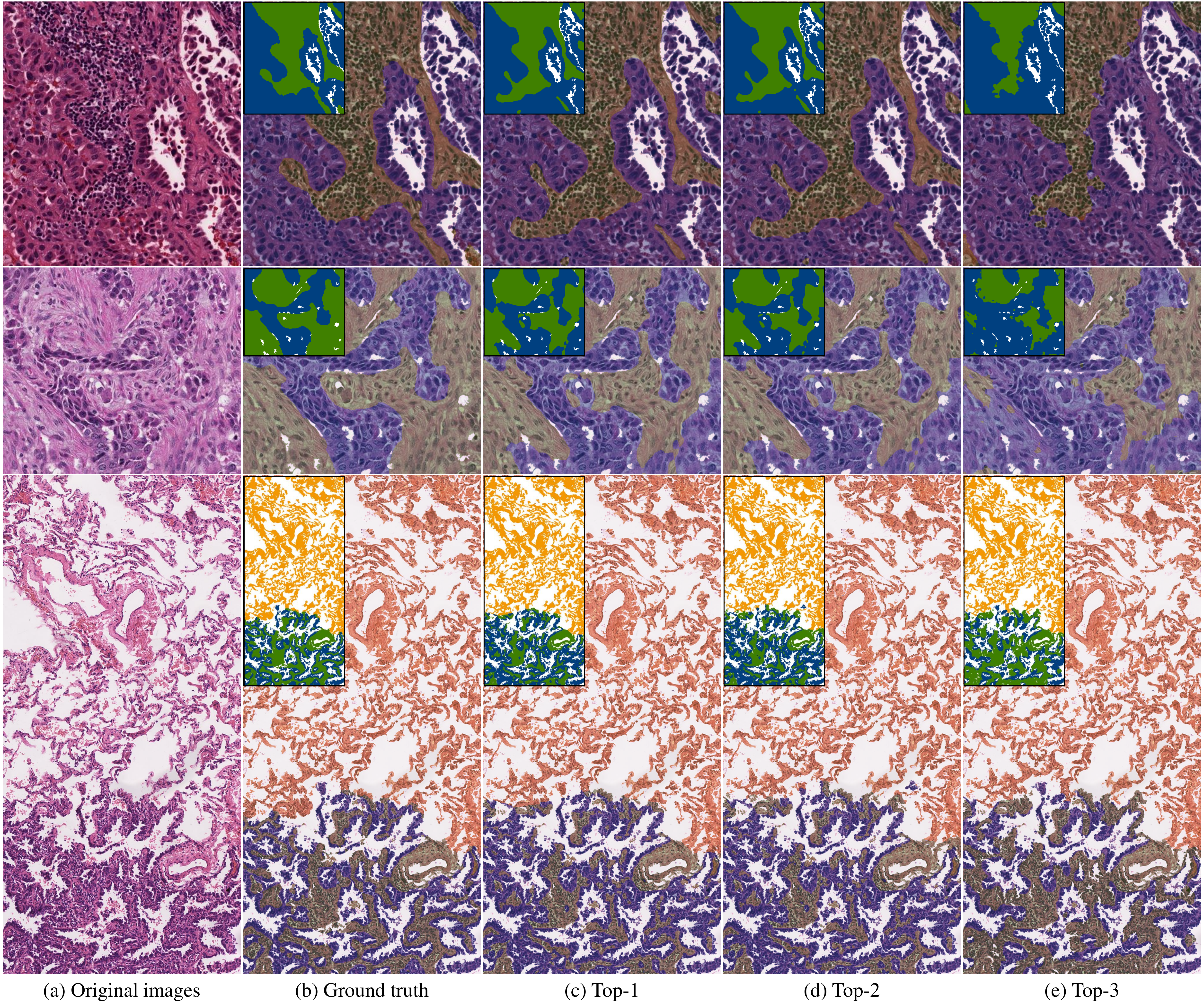}
	\caption{Qualitative results of top-tier approaches. Masks are shown at the top-left corner of the images. For better visualization, we also overlay the semantic segmentation masks on the images.}
	\label{fig:qualitative}
\end{figure*}

\textbf{Model:}
Fang~\etal~use DeepLab V3+ as the segmentation backbone. The model is first pre-trained by the generated samples and the corresponding pixel-level pseudo labels. Then they infer the rest patches with multiple 
categories, and the generated masks are regarded as pseudo masks. Next, the model is trained by the entire dataset. They train multiple models and use an ensemble strategy for the inference.

\begin{figure*}[t]
	\centering
	\includegraphics[width=.99\linewidth]{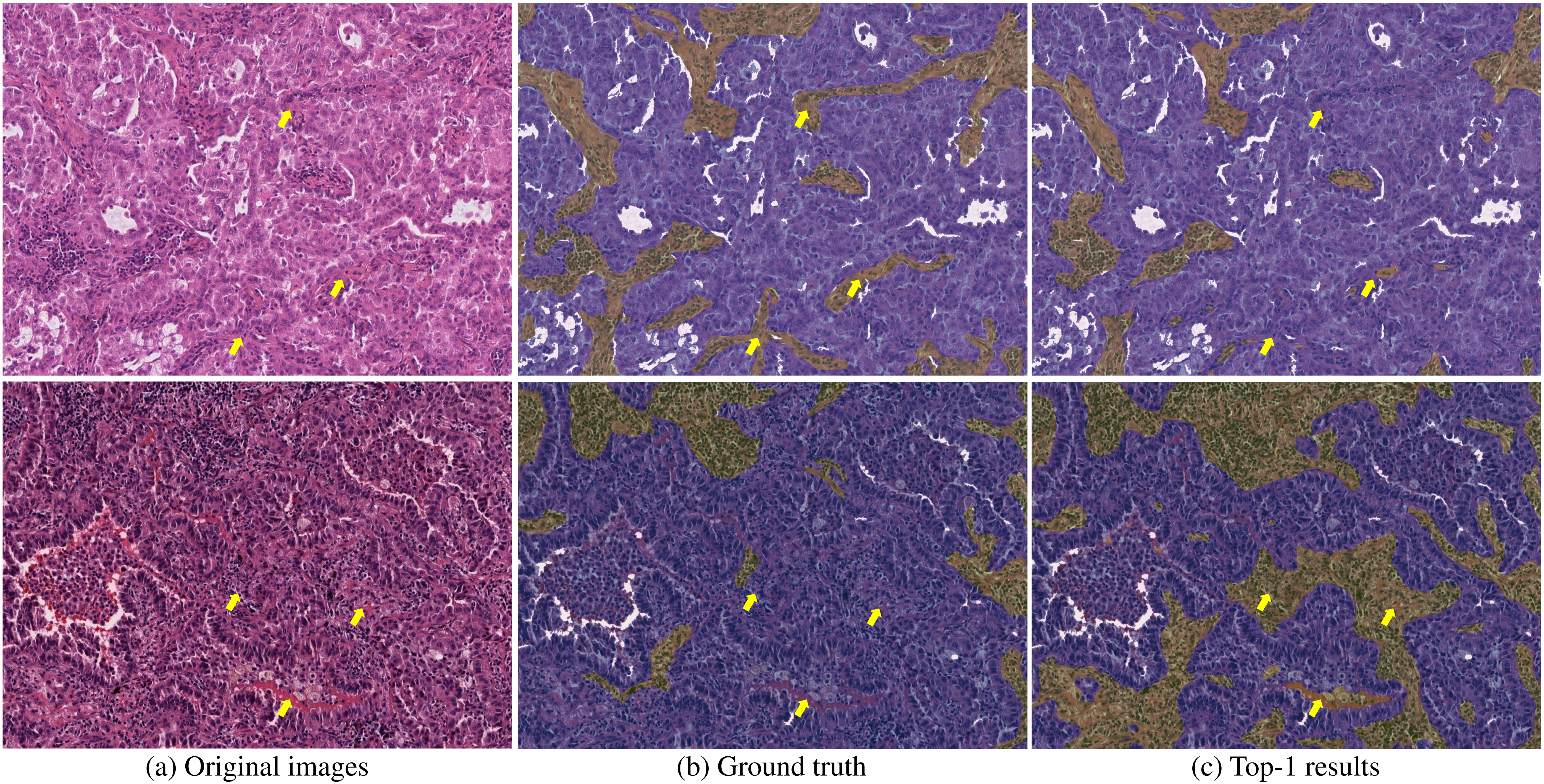}
	\caption{Top-1 model v.s. Ground truth. (a) are two images cropped from the large patches in the test set. (b) are the ground truth masks overlaid on the original images. The yellow arrows highlight some disagreement local areas between GT and prediction results. (c) are the results generated by the Top-1 model overlaid on the original images.}
	\label{fig:qualitative2}
\end{figure*}

\subsubsection{Li~\etal~(Team: shichuanyexi)}
\textbf{Model:}
Li~\etal~proposed an online example mining strategy in the segmentation part to emphasize the credible supervision signals of pseudo-masks from the classification part. And thus,  boost the performance via mitigating the adverse impact of noise in pseudo-masks. Specifically, the loss map $\boldsymbol{L}$ of the segmentation model is normalized by $\frac{\rm{softmax}(- \boldsymbol{L})}{\rm{mean}(\rm{softmax}(- \boldsymbol{L}))}$ to generate the loss weight for weighted cross-entropy loss. From this sampling strategy, easy examples at lower loss values are assigned higher loss weights in optimization dynamically.

Besides this core innovation, this team implemented a two-stage framework for fine-grained predictions based only on the image-level label. It starts from the classification stage, whose backbone is ResNet38~\cite{wu2019wider}. The downsample stride is 8, and the feature maps of the last three stages are fused with interpolations before the classifier for multi-scale representations. All the images are resized into 224 during training. Then the synthesized pseudo-masks are obtained from CAM~\cite{zhou2016learning}. In the second stage, the segmentation framework is PSPNet~\cite{zhao2017pyramid} with backbone ResNet38 at training resolution 224 and testing resolution 320. Data augmentations include random flip, random resized crop, and all patches are normalized by the calculated mean and variance of the competition dataset. Besides, multi-scale testing is applied to both stages.

\subsection{Quantitative and Qualitative Results}
Table~\ref{tab:qualitative} demonstrates the quantitative results of the participants with $mIoU>0.7$. As we can see, the top five teams have achieved over 0.80 mIoU. According to the technical reports in Sec.~\ref{sec:technical}, all the top three teams used the cutmix-like data augmentation way to generate pixel-level labels from single-class patches to provide fully supervision for the segmentation model, which achieved better performance compared with the team that only rely on the patch-level annotations (shichuanyexi). The top-1 team (ChunhuiLin) leverages the prior knowledge that the normal tissue commonly has less crosstalk with tumor epithelial and tumor-associated stroma and achieves the best normal tissue segmentation performance.

In Fig.~\ref{fig:qualitative}, we select three patches (two small patches and a large patch) with different tissues categories, large intra-class varieties and different staining styles to evaluate the robustness of the models. The top-1 team maximizes the values of the entire training labels, including generating pixel-level labels from single-class patches by cutmix and pixel-level pseudo labels from multi-labels patches by CAM. The design of the cross-supervision strategy with the Siamese network can also help suppress the noises in the pseudo masks. Thus they achieve the best qualitative performance. The top-3 team only uses the generated labels from single-class patches and trains a fully-supervised semantic segmentation model. They totally abundant the multi-class patches, which make the model predict less effective and inaccurate boundaries for the patches with tumor epithelial and tumor-associated stroma.

Considering the inherent difficulties of manual annotations on histopathology images, the ground truth is inevitably imperfect with some noises, no matter for pixel-level or for patch-level patches. Therefore, models proposed by the top-tier teams may even perform better than we expected. Fig.~\ref{fig:qualitative2} demonstrates the comparison between the Top-1 model results and ground truth. Some obvious disagreement areas are pointed out by the yellow arrows. We can find that most human experts define the regions from a global perspective. Some very small stroma regions that interact with tumor epithelial are ignored by pathologists. For such a huge patch or even a whole slide image, it is also very difficult to precisely define every single pixel. According to the results generated by the deep learning model shown in Fig.~\ref{fig:qualitative2} (b), the AI model is more sensitive than pathologists to the small regions as long as they have similar features representations. 
\section{Conclusion}
In this paper, we summarized the WSSS4LUAD challenge, a weakly-supervised semantic segmentation challenge for WSIs in lung adenocarcinoma. The original intention of this challenge actually comes from the daily research in computational pathology. For each project, pathologists have to draw a large among of pixel-level labels again and again outside of the working hours. Due to the image specialty, the labeling job cannot be replaced by people with no medical background. Such huge labeling efforts make them hard to focus on the real insight of the research itself. Therefore, we want to leverage novel computer vision techniques to save the annotation efforts.

We invited 11 pathologists to prepare labels for this challenge. A total of 87 WSIs were collected from 87 patients diagnosed with lung adenocarcinoma from both GDPH and TCGA. We introduced a pathologist-in-the-loop strategy to accelerate the labeling process. The final dataset contains 10091 patch-level labels in the training set and over 131 million pixels annotations in the validation and test sets. The entire dataset is released on the challenge website. 

There are 532 registrations from over 40 countries or regions. 28 teams submitted their results with over 1,000 submissions. The best performance team achieves 0.8413 mIoU. According to the technical reports from the top-tier teams, CAM is still the first choice for the WSSS problem. We can see that participants tried their best to minimize the information gap between the patch-level labels and the pixel-level labels in several aspects. In the technical aspect, some teams proposed a multi-task learning strategy with a classification task and a segmentation task to learn more generalizable features. The top-1 team proposed a cross-supervision with Siamese network structure. They used the pseudo masks generated by the Siamese branches to guide each other, which is like an ensemble or voting strategy. In the data augmentation aspect, the cutmix (image splicing) approach has been widely used to introduce more reliable labels by generating fake images from single-class images. Class balancing is also applied to avoid the model being dominated by the majority classes. After the challenge, we also invited the pathologists who were involved in the labeling process to manually assess the segmentation results, which have a high agreement.

From our experience, we believe that annotating patch-level weak labels is a feasible solution to be a complement to the conventional drawing pixel boundaries way for tissue semantic segmentation. It has the great potential to alleviate the annotation efforts. In this challenge, we spent over two months preparing the entire dataset. Most of the time has been spent on preparing the pixel-level validation and test sets, around one and a half months. Preparing the patch-level training set only spent less than a week. Based on the labeling experience in this challenge, we are now refining our pathologist-in-the-loop process with more superior WSSS models to further accelerate the labeling process and improve the segmentation precision.

There are still two limitations to this challenge. First, even we have invited 11 pathologists to prepare the labels coupling with the help of AI models. Imperfect labels are still inevitable, especially pixel-level labels. As shown in Fig.~\ref{fig:qualitative2}, some local results generated by the WSSS model even outperform the ground truth labels. We also observe that the noises in the patch-level dataset are far less than that in the pixel-level dataset. Because the difficulties of these two label ways are totally not on the same level. The above observations lead to another thinking that whether pixel-level labels with more noises are better than patch-level labels with fewer noises. We may follow this thinking and discuss it in our future works. Second, in this challenge, we only selected three majority tissue classes. There are still a lot of minority tissue classes in the WSIs with LUAD, such as necrosis, vessels, bronchus and etc. Some might be also important but only account for an extremely small part of the whole slide image. Further evaluations and discussions should also be included in the future.

\section*{Acknowledgments}
This work was supported by the Key R\&D Program of Guangdong Province, China (No. 2021B0101420006), the National Key R\&D Program of China (No. 2021YFF1201003), the National Science Fund for Distinguished Young Scholars (No.81925023), the National Natural Science Foundation of China (No. 62102103, 81771912, 82071892, 82001789, 81901704, 81702322, 81472712, 81772918, 81972277 and 62002082), High-level Hospital Construction Project (No. DFJH201805 and DFJH201914).

We would like to thank all everyone who contribute the success of WSSS4LUAD challenge.
\textit{Challenge program committee}: Zaiyi Liu, Qingling Zhang, Huihua Yang, Chunming Li, Jun Xu, Zhenbing Liu, Yan Xu, Xipeng Pan, Zhenwei Shi and Chu Han.
\textit{Challenge preparation and technical supports}: Jiatai Lin, Bingchao Zhao, Zhizhen Wang and Yumeng Wang.

We would like to express our sincere thanks to all the pathologists, including the expert pathologist (Lixu Yan), two senior pathologists (Su Yao and Shanshan Lv) and 8 junior labelers (Bingbing Li, Huan Lin, Zeyan Xu, Chao Zhu, Yuan Zhang, Huihui Wang, Huasheng Yao, Jinhai Mai).

\bibliography{ref}
\bibliographystyle{IEEEtran}

\end{document}